\documentclass[aps,fleqn,preprint]{revtex4-1}

\usepackage{amsmath}
\usepackage{graphicx}
\usepackage{epsfig}
\usepackage{subfigure}
\usepackage{amsfonts}
\usepackage{amssymb}
\usepackage{amsthm}
\usepackage{newlfont}
\usepackage{epstopdf}

\usepackage[utf8]{inputenc}

\usepackage{dcolumn}
\usepackage{bm}

\usepackage{xcolor}

\usepackage[section] {placeins}

\begin{document}

\title{Numerical solution to the time-dependent Gross-Pitaevskii equation}

\author{Tsogbayar Tsednee}
\author{Banzragch Tsednee}
\author{Tsookhuu Khinayat}
\affiliation{Institute of Physics and Technology, Mongolian Academy of Sciences, Peace Ave 54B, 13330 Ulaanbaatar, Mongolia }

\begin{abstract}

In this work we employ the split-step technique combined with a Legendre pseudospectral representation to solve various time-dependent Gross-Pitaevskii equations (GPE). Our findings based on the numerical accuracy of this approach applied for one-dimensional (1D) and two-dimensional (2D) problems show that it can provide accurate and stable solutions. Moreover, this approach has been applied to study the dynamics of the Bose-Einstein condensate which is modeled with the GPE. The breathing of condensate with an repulsive and attractive interactions trapped in 1D and 2D harmonic potentials has been simulated as well.

\end{abstract}

\pacs{Valid PACS appear here}
                             

\maketitle

\section{Introduction}

Gross-Pitaevskii theory \cite{Pitaevskii61, Gross61} is one of the successful mathematical models describing the Bose-Einstein condensate (BEC) in which the bosons simultaneously occupy the same quantum state of the lowest energy. The BEC was theoretically predicted by Bose and Einstein \cite{Bose24, Einstein25} in 1925, however, it had first been experimentally obtained in 1995 \cite{Anderson95, Bradley95, Davis95}.

Various methods have been tested to obtain numerical solution for the time-dependent Gross-Pitaevskii equation (GPE). Weideman~\cite{Weidiman86} employed a split-step Fourier pseudospectral method. Adhikari~\cite{Muruganandam03} used a split-step Crank-Nicolson approach for solving the GPE. Moreover, Bao~\cite{Bao03} and Wang~\cite{Wang05, Wang16} have made a lot of contributions for developing numerical techniques for solving a single and coupled GPEs with different trapping potentials. 

In this work our goal is to solve the time-dependent GPE using the split-step technique combined with a Legendre pseudospectral representation. To our knowledge, the split-step Legendre pseudospectral approach has not been applied to this problem in a systematic and detailed way, yet. In our calculation, we will initially test a numerical accuracy of the proposed approach for a nonlinear one-dimensional (1D) and two-dimensional (2D) Schr\"{o}dinger equation for which an analytical solution is known. Then we will apply the approach for solving the time-dependent GPE which simulates the dynamics of the BEC with a repulsive and an attractive interactions trapped in 1D and 2D harmonic potentials at zero temperature. 

The paper is organized as follows. In Section 2, we present the formalism of the Gross-Pitaevskii theory for the harmonic trap. In Section 3 we give a brief discussion of a split-step technique and a pseudospectral method for the 1D and 2D problems. In Section 4 we present the numerical results and discussions for test problems and the dynamics of condensate trapped in 1D and 2D harmonic potentials. Then a conclusion follows.  

\section{Gross-Pitaevskii theory for trapped bosons}

The dynamics of boson particles confined in the trap (potential) can be modeled with the following time-dependent Gross-Pitaevskii equation \cite{Pitaevskii61, Gross61, Bao03, Wang05, Wang16}:    	
\begin{eqnarray}\label{gpe_3d}
	i\hbar \frac{\partial \psi (\mathbf{r}, t)}{\partial t} = \Big( -\frac{\hbar^{2}}{2m} \nabla^{2}_{\mathbf{r}} + V + \frac{4\pi \hbar^{2} a_{s} N}{m} |\psi(\mathbf{r},t)|^{2} \Big) \psi(\mathbf{r}, t).
\end{eqnarray}	
Here $\psi(\mathbf{r}, t)$ is the BEC wave function, $m$ is the mass of boson, $V$ is an external confining potential (trap), $a_{s}$ is the $s$-wave scattering length and $N$ is the number of bosons in the condensate defined as $N \equiv \langle \psi(\mathbf{r}, t) | \psi(\mathbf{r}, t) \rangle$, and the energy per particle is defined as 
\begin{eqnarray}\label{En_f}
	E(\psi) = \int d\mathbf{r}\Big[\frac{\hbar^{2}}{2 m}|\nabla_{\mathbf{r}} \psi(\mathbf{r}, t)|^{2} + V|\psi(\mathbf{r}, t)|^{2} + \frac{4\pi \hbar^{2} a_{s} N}{m} |\psi(\mathbf{r})|^{4}  \Big].  
\end{eqnarray}	
Usually the external trap potential $V$ is the harmonic trapping potential $\frac{m \omega^{2}_{x} x^{2}}{2} + \frac{m \omega^{2}_{y} y^{2}}{2} + \frac{m \omega^{2}_{z} z^{2}}{2}$
In the trapping potential $\omega_{x}, \omega_{y}$ and $\omega_{z}$ are the frequencies along the $x$-, $y$- and $z$-directions, respectively. 

It is convenient to use the equation (\ref{gpe_3d}) in the dimensionless form in the calculation. By choosing the $\tilde{t} = \omega_{x} t$ and $a_{x} = (\hbar/ m \omega_{x})^{1/2}$ as the dimensionless time and length units, respectively, we can have the rescaled quantities: $\tilde{\mathbf{r}} = \mathbf{r}/a_{x}$, $\tilde{\psi}(\tilde{\mathbf{r}},\tilde{t}) = a^{3/2}_{x}/N^{1/2}\psi(\mathbf{r},t)$, and $\tilde{E} =  E/\hbar \omega_{x}$. After removing all $\tilde{}$, the time-dependent GPE (\ref{gpe_3d}) can be written in the dimensionless form \cite{Bao03, Wang05, Wang16}:
\begin{eqnarray}\label{gpe_3d_2}
	i\frac{\partial \psi (\mathbf{r}, t)}{\partial t} = \Big[ -\frac{1}{2} \nabla^{2}_{\mathbf{r}} + V + g |\psi(\mathbf{r},t)|^{2} \Big] \psi(\mathbf{r}, t),
\end{eqnarray}  
where the nonlinear constant $g=4\pi a_{s} N/a_{x}$. In this work we use the 1D and 2D harmonic trapping potentials given in the Cartesian coordinates. In 1D, $\mathbf{r} = (x)$, and the trapping potential $V$ in equation (\ref{gpe_3d_2}) is  $V(x) = \frac{x^2}{2}$. For the 2D case, $\mathbf{r} = (x,y)$ and $V(x,y) = \frac{x^{2}}{2} + \frac{\lambda^{2} y^{2}}{2}$, with $\lambda = \frac{\omega_{y}}{\omega_{x}}$.

We note that the equation (\ref{gpe_3d_2}) has no analytic solution, however, it can be solved employing the numerical techniques which I will discuss in the next section.

\section{Numerical procedure} 

In this work we consider consider the initial-boundary value problem for the GPE given in equation (\ref{gpe_3d_2}). We can write it in the form:
\begin{eqnarray}\label{gpe_cn_1}
	i\frac{\partial \psi (\mathbf{r}, t)}{\partial t} = H\psi(\mathbf{r}, t)  
\end{eqnarray}
with an initial condition $\psi(\mathbf{r}, 0) = \psi_{0}(\mathbf{r}), \quad \mathbf{r}\in \mathbb{R}^{d}\, (d=1,2,3)$, and a boundary condition $\lim\limits_{|\mathbf{r}|\to \pm \infty} \psi(\mathbf{r}, t) = 0, \quad t\geq 0$. Here $d$ means a dimension of space.
A formal solution for this equation can be written as
\begin{eqnarray}\label{gpe_cn_1_sol}
	\psi (\mathbf{r}, t+ \Delta t) = e^{-i H \Delta t} \psi(\mathbf{r}, t). 
\end{eqnarray}

In our numerical calculation we solve the equation (\ref{gpe_cn_1}) for two cases: (i) without a split-step technique and (ii) with a split-step technique. In both cases when we solve the equation (\ref{gpe_cn_1}), we employ the Crank-Nicolson (CN) scheme in time which is unconditionally stable and the Legendre-pseudospectral method in space part of the Hamiltonian $H$. 

For the first case in which a split-step technique is not used a solution for equation (\ref{gpe_cn_1}) is indeed given in the form (\ref{gpe_cn_1_sol}), and can be found with the CN scheme:
\begin{eqnarray}\label{gpe_cn_1_sol_wo}
	\psi^{n+1} = \Big(1 + \frac{i \Delta t}{2} H \Big) ^{-1}\Big(1 - \frac{i \Delta t}{2} H \Big) \psi^{n}, 
\end{eqnarray} 
where $\psi^{n} = \psi(\mathbf{r}, t_{n})$ and $\Delta t >0$ is a time step size with which a time step is given as $t_{n} = n \Delta t, \, n=0, 1, \ldots$

Now let's consider the second case in which the split-step technique is employed. The main idea of this approach is that we divide the Hamiltonian operator $H$ into two parts: a linear part $A = -\frac{1}{2}\nabla^{2}_{\mathbf{r}}$ and a non-linear part $B = V + g |\psi(\mathbf{r},t)|^{2}$. Then we solve the  equation (\ref{gpe_cn_1}) in terms of two successive steps. In the Step 1, we solve the following a nonlinear operator equation 
\begin{eqnarray}\label{gpe_step1}
	i\frac{\partial \psi (\mathbf{r}, t)}{\partial t} = B\psi(\mathbf{r}, t)
\end{eqnarray}  
for the time step $\Delta t$; then Step 2, followed by solving 
\begin{eqnarray}\label{gpe_step2}
	i\frac{\partial \psi (\mathbf{r}, t)}{\partial t} = A\psi(\mathbf{r}, t)
\end{eqnarray} 
for the same time step. 

From the nonlinear subproblem (\ref{gpe_step1}) in the Step 1, for $t_{n} \leq t \leq t_{n+1}$ we can have an exact solution in the form
\begin{eqnarray}\label{gpe_step1_sol}
	\psi(\mathbf{r}, t) = e^{-i B \Delta t } \psi(\mathbf{r}, t_{n}) = e^{-i (V(\mathbf{r}) + g |\psi(\mathbf{r}, t_{n})|^{2}) \Delta t } \psi(\mathbf{r}, t_{n}). 
\end{eqnarray}  
In Step 2, we obtain a numerical solution in the following form:
\begin{eqnarray}\label{gpe_step2_sol}
	\psi^{n+1} =  \Big( 1 + \frac{i \Delta t}{2} A\Big)^{-1}\Big( 1 - \frac{i \Delta t}{2} A\Big)\psi^{n},
\end{eqnarray} 
which is an exactly same approach as given in (\ref{gpe_cn_1_sol_wo}). Then the solution (\ref{gpe_cn_1_sol})  for the equation can be found in the form:
\begin{eqnarray}\label{gpe_step12_sol}
	\psi(\mathbf{r}, t+\Delta t) = e^{-i A\Delta t} e^{-i B \Delta t}\psi(\mathbf{r},t).
\end{eqnarray} 
We note that this splitting technique has only first-order in time (equation (\ref{gpe_step12_sol})), so that it can easily be implemented. Note that one may use higher-order splitting approaches  \cite{Wang05, Wang16, Bandrauk93} to solve the GPE as well.  

For numerical calculation of $\nabla^{2}_{\mathbf{r}}$ in $H$  and $A$ operators, we used the Legendre differentiation matrix of the second order ($d^{2}_{ij}$) (details can be found in \cite{Tsogbayar13, Tsogbayar21}). For 1D case, $\nabla^{2}_{xx} \approx  d^{2}_{xx}$. For 2D case, $d^{2}_{xx}\otimes I_{yy} + I_{xx}\otimes d^{2}_{yy}$, where $I$ is the unit matrix and $\otimes$ denotes the Kronecker product \cite{Tsogbayar13, Tsogbayar21}. We note that this Legendre pseudospectral approach had been employed to obtain a solution for the time-independent GPE for an anisotropic 3D trapping potential in our previous work~\cite{Tsogbayar21}.    	

Below we have shown the solutions (\ref{gpe_cn_1_sol_wo}) and (\ref{gpe_step12_sol}) in the discrete forms which can directly be implemented in coding.  For the 1D GPE, the solution (\ref{gpe_cn_1_sol_wo}) can be written as
\begin{eqnarray}\label{gpe_step1_sol_disc}
	\psi^{n+1}_{i}  & = & \Big[ I + \frac{i \Delta t}{2} \Big(-\frac{1}{2}d^{2}_{xx} +  \mathrm{diag}[V_{i}] + g \, \mathrm{diag}[|\psi^{n}_{i}|^{2}]\Big)\Big]^{-1} \\ \nonumber
	 & & \times  \Big[  I - \frac{i \Delta t}{2} \Big(-\frac{1}{2}d^{2}_{xx} +  \mathrm{diag}[V_{i}] + g\, \mathrm{diag}[|\psi^{n}_{i}|^{2}]\Big) \Big] \psi^{n}_{i}. 	
\end{eqnarray} 
The  solution (\ref{gpe_step12_sol}) can be obtained as 
\begin{eqnarray}\label{gpe_step2_sol_disc}
	\psi^{n+1}_{i} & = &   \Big[ I - \frac{i \Delta t}{4} d^{2}_{xx} \Big]^{-1} \Big[  I + \frac{i \Delta t}{4} d^{2}_{xx}  \Big] 
	 \exp[ -\Delta t(V_{i} + g\, |\psi^{n}_{i}|^{2} )]  \psi^{n}_{i}. 
\end{eqnarray}
In both (\ref{gpe_step1_sol_disc}) and (\ref{gpe_step2_sol_disc}) equations, we use the notations: $\psi^{n}_{i} \equiv \psi (x_{i}, t_{n})$, $I$ is unit matrix, ``$\mathrm{diag}$" means a diagonal matrix, and $x_{i} $ are non-uniformly distributed $N_{x}+1$ number of the Legendre-Gauss-Lobatto grid points in an interval $[a, b]$.  Note that in numerical calculation we need to take into account the boundary condition in such a way that two end points of a column vector will not be used for the Dirichlet boundary condition, and for the 2D case in discretized form a main difference comes with the Kronecker product as mentioned above \cite{Tsogbayar21}.

\section{Results and discussion} 

\subsection{Numerical tests}

In our numerical experiments we initially test the numerical accuracy of two approaches: (i) the ordinary Legendre-pseudospectral (LS) and (ii) the split-step Legendre-pseudospectral (SSLS) methods through the following two examples. 

{\bf Example 1.} First we consider the following GPE in 1D \cite{Wang05} 
\begin{eqnarray}\label{num_test_1d}
	i \psi_{t} = -\psi_{xx} - 2 |\psi|^{2}\psi
\end{eqnarray}
with the homogeneous Dirichlet condition over $[-20, 20]$.  The analytical solution for this equation is $\psi_{\mathrm{exact}}(x,t) = sech(x-4t)e^{i(2x-3t)}$. This is the usual non-linear Schr\"{o}dinger equation since it does not include the trapping potential $V$ (GPE (\ref{gpe_3d})). We computed the maximum error between the exact solution and the approximated solution which are obtained from the LS and the SSLS methods. Table 1 shows the maximum of $|\psi_{\mathrm{exact}}(x,t) - \psi_{\mathrm{approx}}(x,t) |$ until $t=4$. A number of grid points along $x$-axis is $N_{x}=128$ and the time step size $\Delta t = 0.01$.  We note that this problem was also considered using the split-step finite-difference (SSFD) and split-step Fourier spectral (SSFS) methods in Ref.~\cite{Wang05}, and at $t=4$ with $\Delta t = 0.01$ value of this error obtained by the SSFD approach is $3.6620\cdot 10^{-2}$, while the SSFS method's estimation for it is $3.663\cdot 10^{-2}$ (Table 1 in \cite{Wang05}). In Ref.~\cite{Wang05} the spatial mesh size $\Delta x = 0.01 = (b-a)/M, a = -20, b = 20$, where $M$ is number of grid points. Then a number of grid point used in \cite{Wang05} is $M = 4000$.  This is quite large number compared with the number we used ($N_{x} = 128$), indeed. We say that a reason why our calculation with relatively few number of grid points works well is that the Legendre-Gauss-Lobatto grid points are distributed non-uniformly in the interval, that is, more points come near two ends of the interval \cite{Tsogbayar17}. 

\begin{table}[h]
	\caption{ The maximum error between the exact solution and approximated solution at different time $t$ in Example 1. $N_{x} = 128$ and $\Delta t = 0.01$.  }
	\begin{center}
		{\scriptsize
			\begin{tabular}{c@{\hspace{2mm}}c@{\hspace{2mm}}c@{\hspace{2mm}}c@{\hspace{2mm}}c@{\hspace{2mm}}c@{\hspace{2mm}}c@{\hspace{2mm}}c@{\hspace{2mm}}c@{\hspace{2mm}}c@{\hspace{2mm}}c@{\hspace{2mm}}c@{\hspace{2mm}}c@{\hspace{2mm}}c@{\hspace{2mm}} }
				\hline\hline
				$t$ & LS (Eq.~(\ref{gpe_cn_1_sol_wo}))  & SSLS (Eq.~(\ref{gpe_step12_sol})) \\
				\hline
				0.5 & 1.1e-2 & 4.7e-3 \\
				1 & 2.4e-2 & 6.6e-3\\
				2 & 5.7e-2 & 1.1e-2 \\ 
				3 & 1.1e-1 & 1.5e-2 \\
				4 & 1.8e-1 & 3.6e-2 \\ 
				\hline\hline
		\end{tabular} }
	\end{center}
\end{table}

{\bf Example 2.} We consider the following GPE in 2D \cite{Wang05} 
\begin{eqnarray}\label{num_test_2d}
	i\psi_{t} =  -\frac{1}{2} (\psi_{xx} + \psi_{yy}) + V(x,y) \psi + |\psi|^{2}\psi, \quad (x,y)\in [0, 2\pi] \times [0,2\pi], 
\end{eqnarray}
where $V(x,y) = 1 - \sin^{2}x\sin^{2}y$.  An initial condition $\psi_{0}(x,y) = \sin x \sin y$ are used. The exact solution for this problem with the Dirichlet boundary condition is $\psi_{\mathrm{exact}}(x,y,t) = \sin x \sin y e^{-i 2t}$. Table 2 presents the maximum of $|\psi_{\mathrm{exact}}(x,y,t) - \psi_{\mathrm{approx}}(x,y,t)|$ until time $t=30$ using the LS and SSLS methods. The number of grid points along the $x$- and $y$-axis is $N_{x} = N_{y} = 16$ and the time step size $\Delta t = 0.01$. This example problem was also discussed in Ref.~\cite{Wang05} where an author used the SSFD and SSFS methods. The maximum errors from the SSFD and SSFS methods with $N_{x} = N_{y} = 128$ at $t=20$ with $\Delta t = 0.01$ are $4.057\cdot 10^{-3}$ and $1.138\cdot 10^{-10}$, respectively \cite{Wang05}. From the errors estimated, we see that the SSFS approach has shown an extremely good result since its solution can be obtained with an almost analytical expression in terms of the ordinary Fourier pseudospectral representation. 

\begin{table}[h]
	\caption{ The maximum error between the exact solution and approximated solution at different time $t$ in Example 2.  $N_{x} = N_{y} = 16$ and $\Delta t = 0.01$.  }
	\begin{center}
		{\scriptsize
			\begin{tabular}{c@{\hspace{2mm}}c@{\hspace{2mm}}c@{\hspace{2mm}}c@{\hspace{2mm}}c@{\hspace{2mm}}c@{\hspace{2mm}}c@{\hspace{2mm}}c@{\hspace{2mm}}c@{\hspace{2mm}}c@{\hspace{2mm}}c@{\hspace{2mm}}c@{\hspace{2mm}}c@{\hspace{2mm}}c@{\hspace{2mm}} }
				\hline\hline
				$t$ & LS (Eq.~(\ref{gpe_cn_1_sol_wo}))  & SSLS (Eq.~(\ref{gpe_step12_sol})) \\
				\hline
				1 & 6.6e-5 & 8.2e-6 \\
				5 & 3.3e-4 & 4.1e-5 \\ 
				10 & 6.6e-4 & 8.2e-5 \\
				20 & 1.3e-3 & 1.6e-4 \\ 
				30 & 2.0e-3 & 2.5e-4 \\
				\hline\hline
		\end{tabular} }
	\end{center}
\end{table}

From Tables 1 and 2, we have seen that results (3rd column in each table) from the SSLS method are more accurate than those (2nd column in each table) from the ordinary LS approach. 
We assume that a reason for this can be related to a fact that the SSLS approach uses the exact approximation in an intermediate step (Step 1).  
Thus, we will employ the split-step technique in the succeeding numerical calculations.  Moreover, we want to note that our LS and SSLS approaches can be considered as applicable ones 
 even for small number of grid points compared to other approaches, such as the SSFS, used in \cite{Wang05}. We also note that more accurate results from the LS and SSLS approaches,  especially for large computational domains of time and space can be easily obtained by handling the computational parameters, such as, increasing number of grid points.

\subsection{Numerical applications}

In this subsection based on numerical experiences obtained in a preceding subsection we will solve the GPE (\ref{gpe_3d_2}) of which an analytical solution is unknown. We will apply the split-step technique for the following examples which have a trap potential.

{\bf Example 3.} We solve the GPE (\ref{gpe_3d_2}) in 1D 
\begin{eqnarray}\label{num_test_1d}
	i \psi_{t} = -\frac{1}{2}\psi_{xx} + \frac{x^{2}}{2} + g |\psi|^{2}\psi 
\end{eqnarray}
with an initial condition $\psi_{0}(x) = \pi^{-1/4}e^{-x^{2}/2}$, which is a ground state for the trap with $g=0$. 
When $g>0$, the interaction between particles is repulsive. In contrast, when $g<0$, the interaction is attractive. 
We solve this problem on $[-10, 10]$ with a Dirichlet boundary condition. 

In Figure 1a we have shown the time-evolution of the condensate $|\psi(x,t)|^{2}$ for a repulsive interaction ($g=5$). 
The dynamics of a ground state (an initial state) that is left to evolve with a different trapping potential is called a ``breathing'' \cite{Dion03, Ruprecht95}. 
The ``breathing'' of the condensate shown in panel 1a may happen because the initial trapping frequency is reduced, resulting in expansion of the trapping potential \cite{Dion03}. 
Panel 1b of Figure 1 presents how the nonlinear constant $g$ affects the breathing frequency of the condensate. It's been clearly shown that the density at the center of the trap is lower for a  larger value of $g$, which is expected because of the corresponding higher 
inter-particle repulsion.     
%
\begin{figure}[ht]
	\centering
	\mbox{\subfigure[]{\includegraphics[width=0.4200\textwidth]{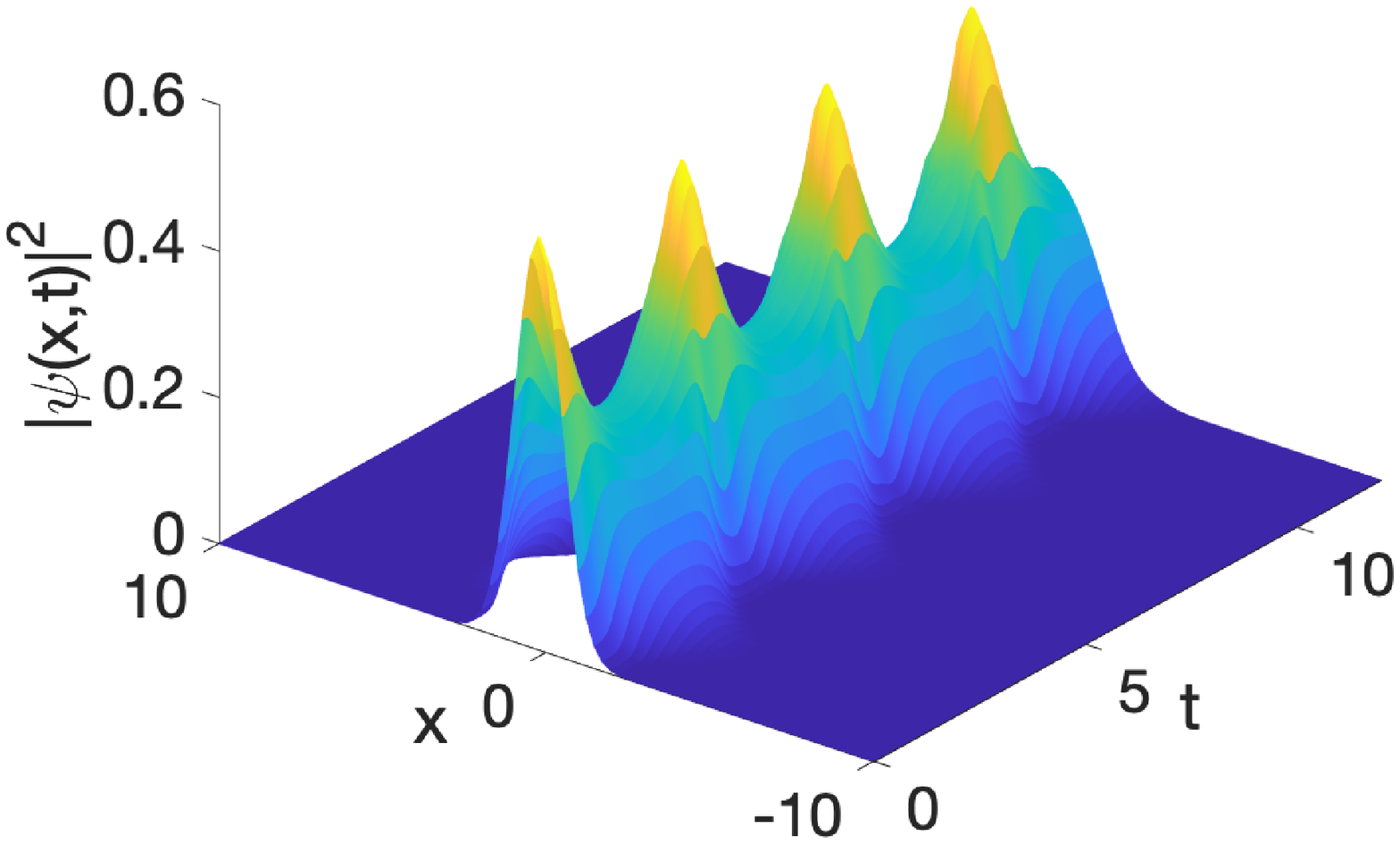}}
		\subfigure[]{\includegraphics[width=0.4200\textwidth]{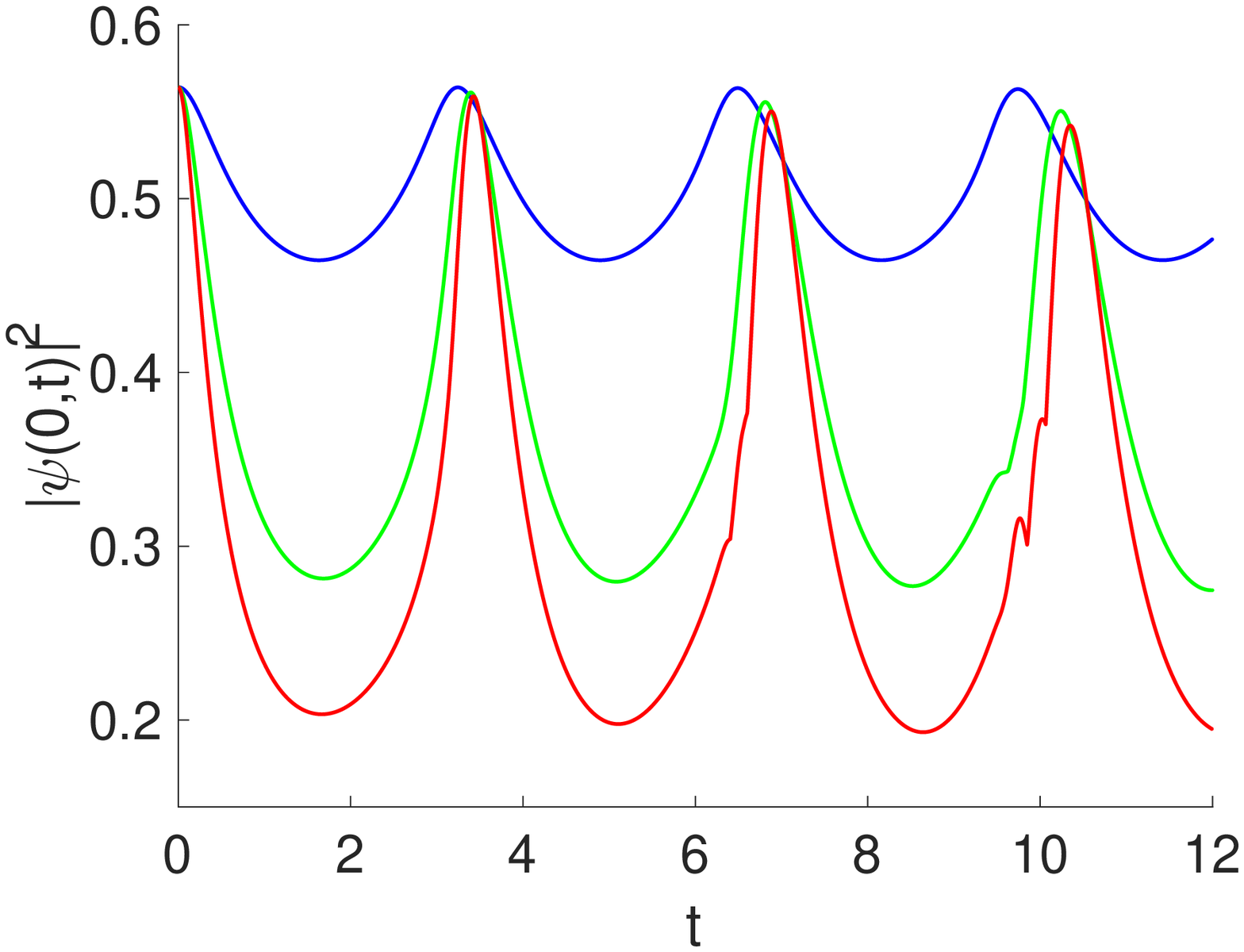}}
	}
	\caption{(a) ``Breathing'' of the condensate for the repulsive interaction with $g=5$. (b) The value of the wave function  $|\psi (0,t)|^{2}$ in the center of the trap is given as a function of time $t$. Blue, green and red curves correspond to values of the nonlinear constant $g=1$, $5$ and $10$, respectively. Computational parameters: $N_{x} = 128$, $\Delta t = 0.001$.       } 
\end{figure}

Figure 2a presents the dynamics of the condensate for which an inter-particle interaction is attractive ($g=-2.5$). Because of this attractive behavior, density at the center of the trap $|\psi(0,t)|^{2}$ is higher for a smaller value of $g$ (Panel 2b). We also observe that the oscillation frequency of the condensate in the trap increases with a smaller value of $g$ since more attractive potential energy is added.      

\begin{figure}[ht]
	\centering
	\mbox{\subfigure[]{\includegraphics[width=0.4200\textwidth]{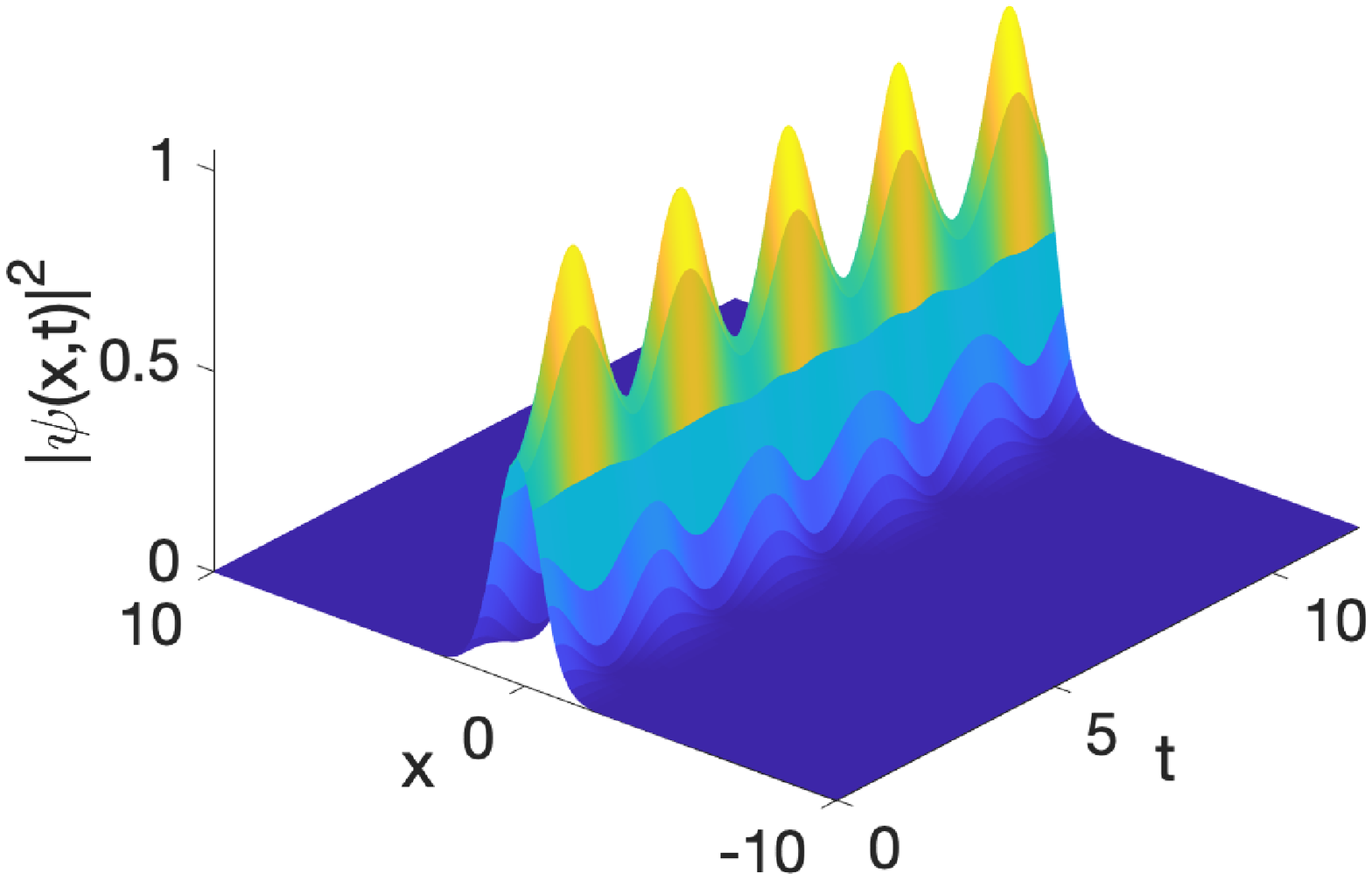}}
		\subfigure[]{\includegraphics[width=0.4200\textwidth]{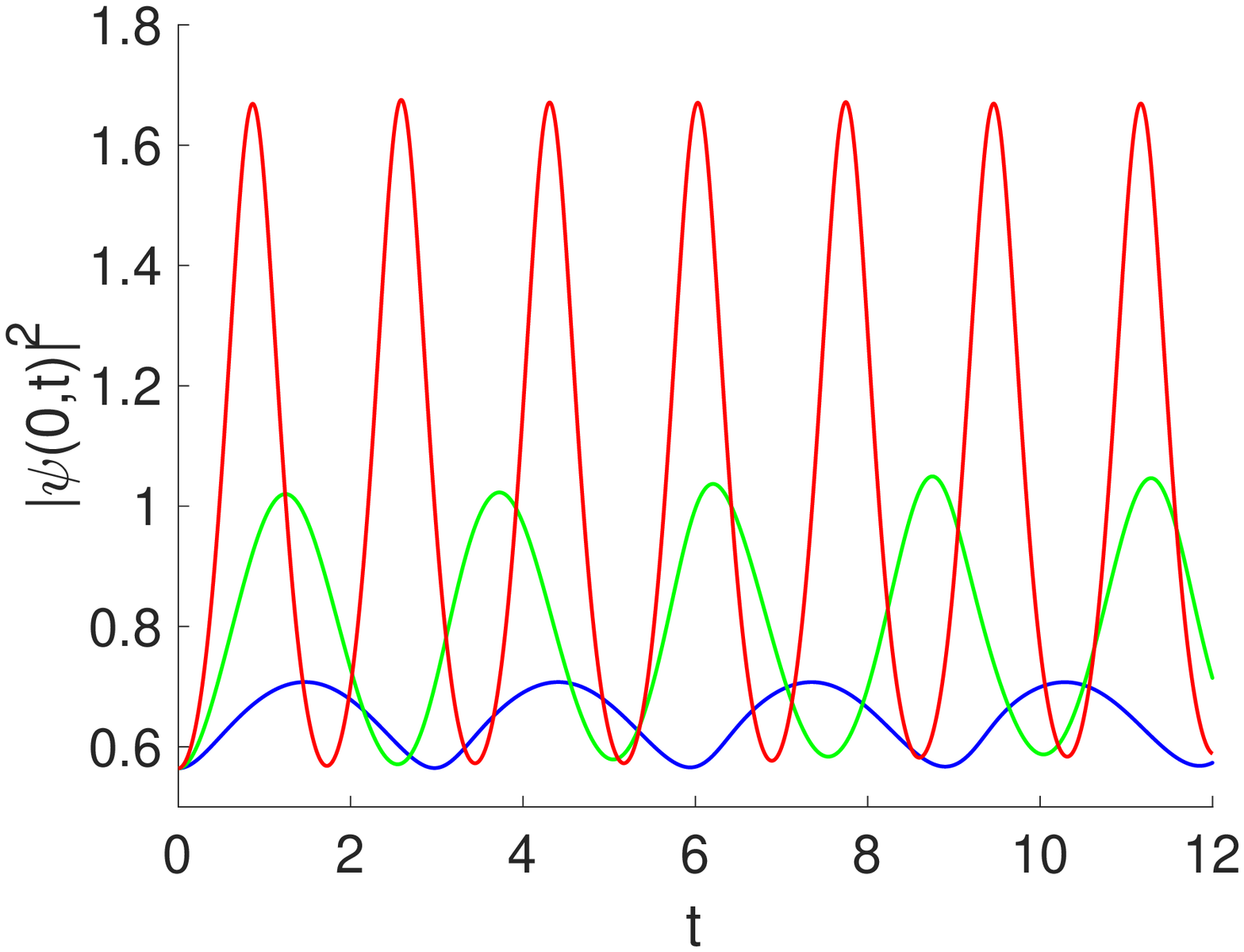}}
	}
	\caption{Same plots as shown in Figure 1, but for $g<0$: (a) the nonlinear constant $g=-2.5$. (b) Blue, green and red curves correspond to $g=-1$, $-2.5$ and $-10$, respectively.  } 
\end{figure}

\begin{figure}[ht]
	\centering
	\mbox{\subfigure[]{\includegraphics[width=0.4200\textwidth]{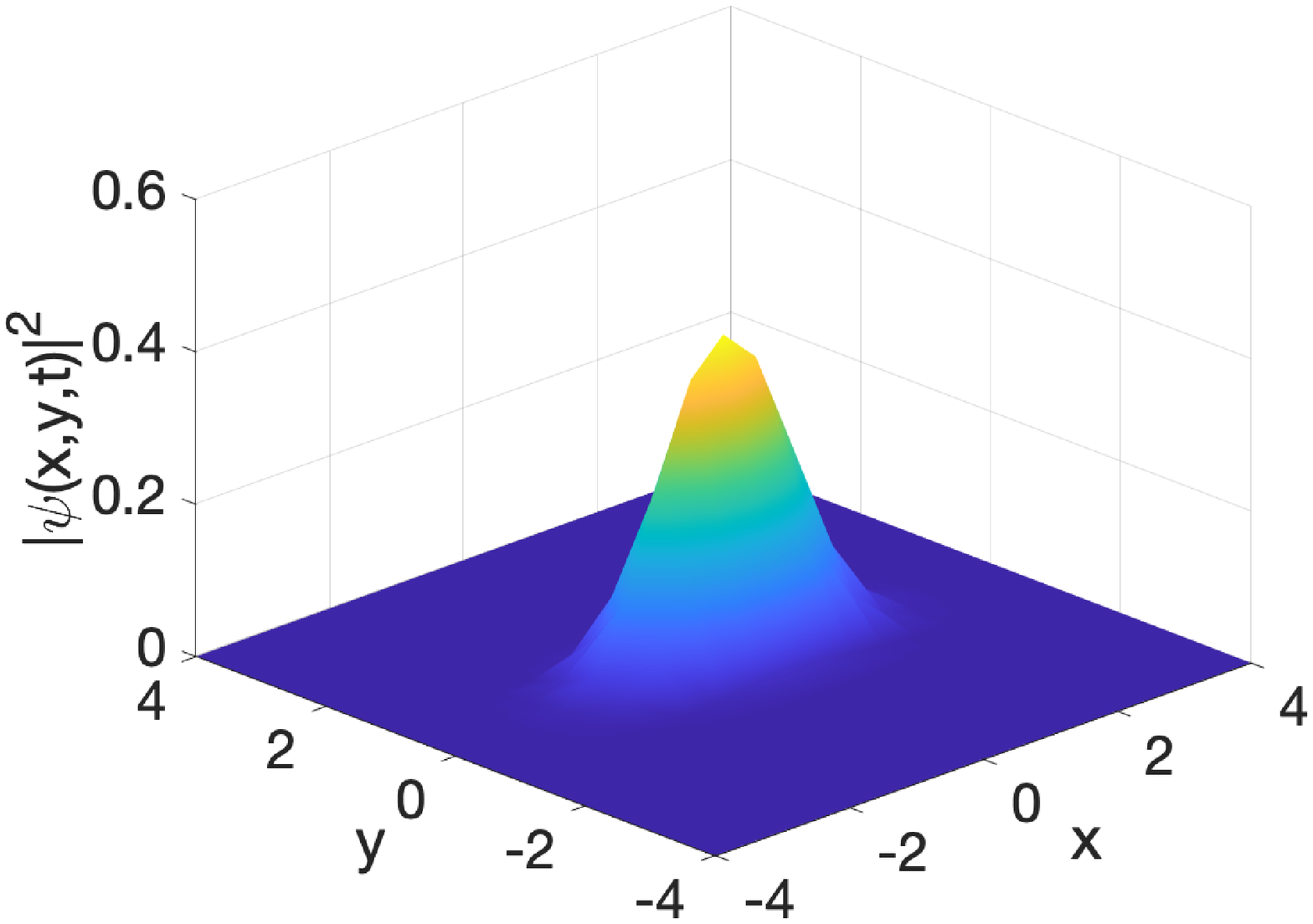}}
		\subfigure[]{\includegraphics[width=0.4200\textwidth]{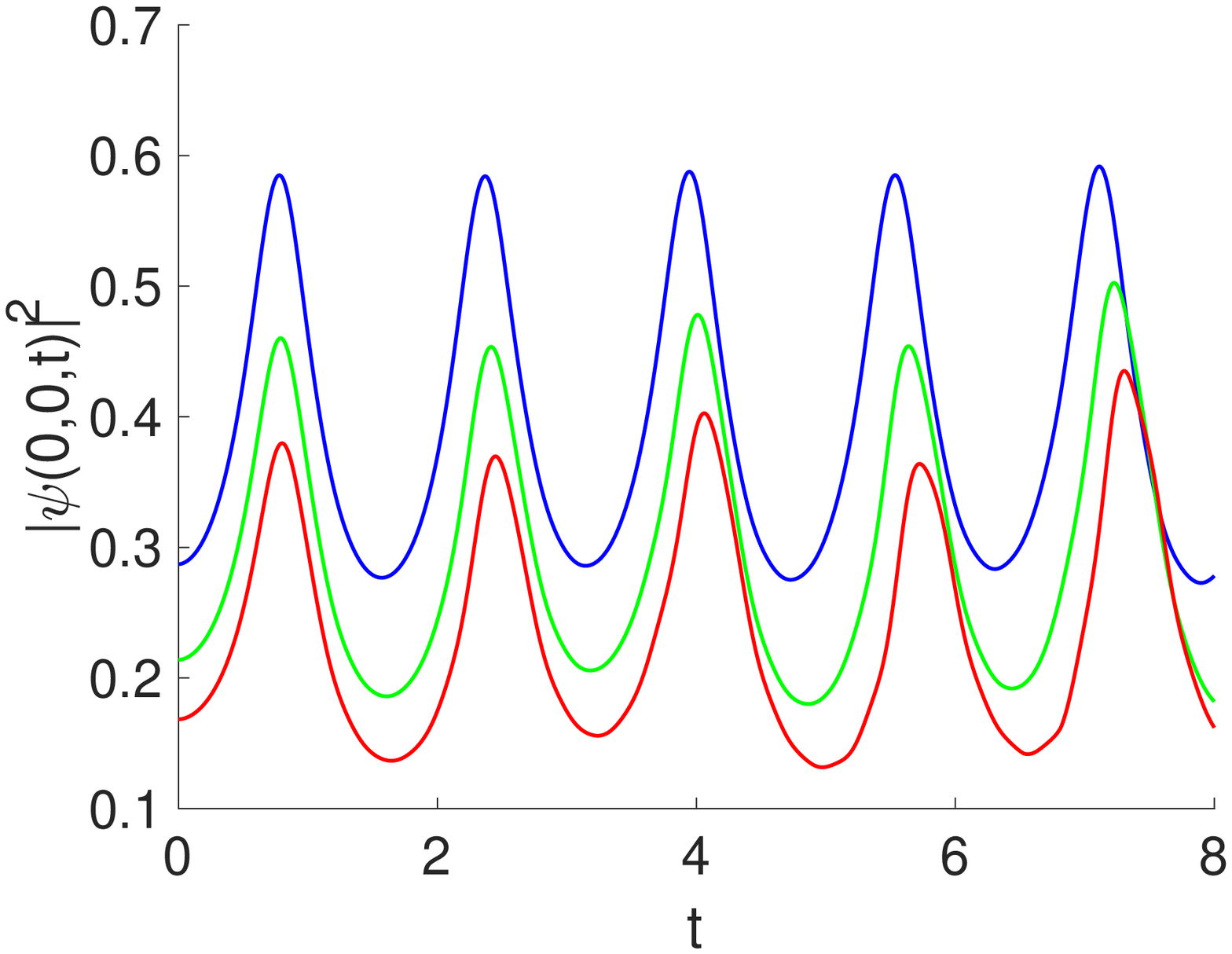}}
	}
	\caption{Surface plot of density $|\psi(x,y,t)|^{2}$ at time $t=2.5$ and $g=5$. (b) ``Breathing'' of the condensate after the isotropic trap becomes the anisotropic trap. The value of $|\psi(x,y,0)|^{2}$ in the center of the trap is given as a function of time $t$ for $g$ equal to $1$ (blue curve), $5$ (green curve) and $10$ (red curve). Numerical parameters: $N_{x}=N_{y} = 86$ and $\Delta t = 0.01$. } 
\end{figure}

In the above two calculations, we chose the number of grid points along $x$-axis $N_{x} = 128$ and the time step size $\Delta t = 0.001$. For the calculations the initial condition $\psi_{0}(x)$ is used, therefore all curves shown in panels 1b and 2b have the same starting point. We note that all calculations are independent on the numerical parameters, since a calculated discrete error norm $||\psi_{N_{x}=128}(t) - \psi_{N_{x}=64}(t)||_{l^{2}}$ at time $t=12$ is $10^{-6}$.    

{\bf Example 4.} We solve the GPE (\ref{gpe_3d_2}) in anisotropic 2D trap 
\begin{eqnarray}\label{num_test_1d}
	i \psi_{t} = -\frac{1}{2}(\psi_{xx}+ \psi_{yy}) + \frac{x^{2}}{2} + \frac{\lambda^{2}y^{2}}{2} + g |\psi|^{2}\psi. 
\end{eqnarray}
The initial condition for this problem $\psi_{0}(x,y)$ is taken as the ground state for an isotropic trap ($\lambda^{2} = 1$). Then we let this initial state $\psi_{0}$ evolve in an anisotropic trap with $\lambda^{2} = 4$. We solve this problem on $[-10, 10]^{2}$ and a number of grid points $N_{x} = N_{y} = 86$.

\begin{figure}[ht]
	\centering
	\mbox{\subfigure[]{\includegraphics[width=0.4200\textwidth]{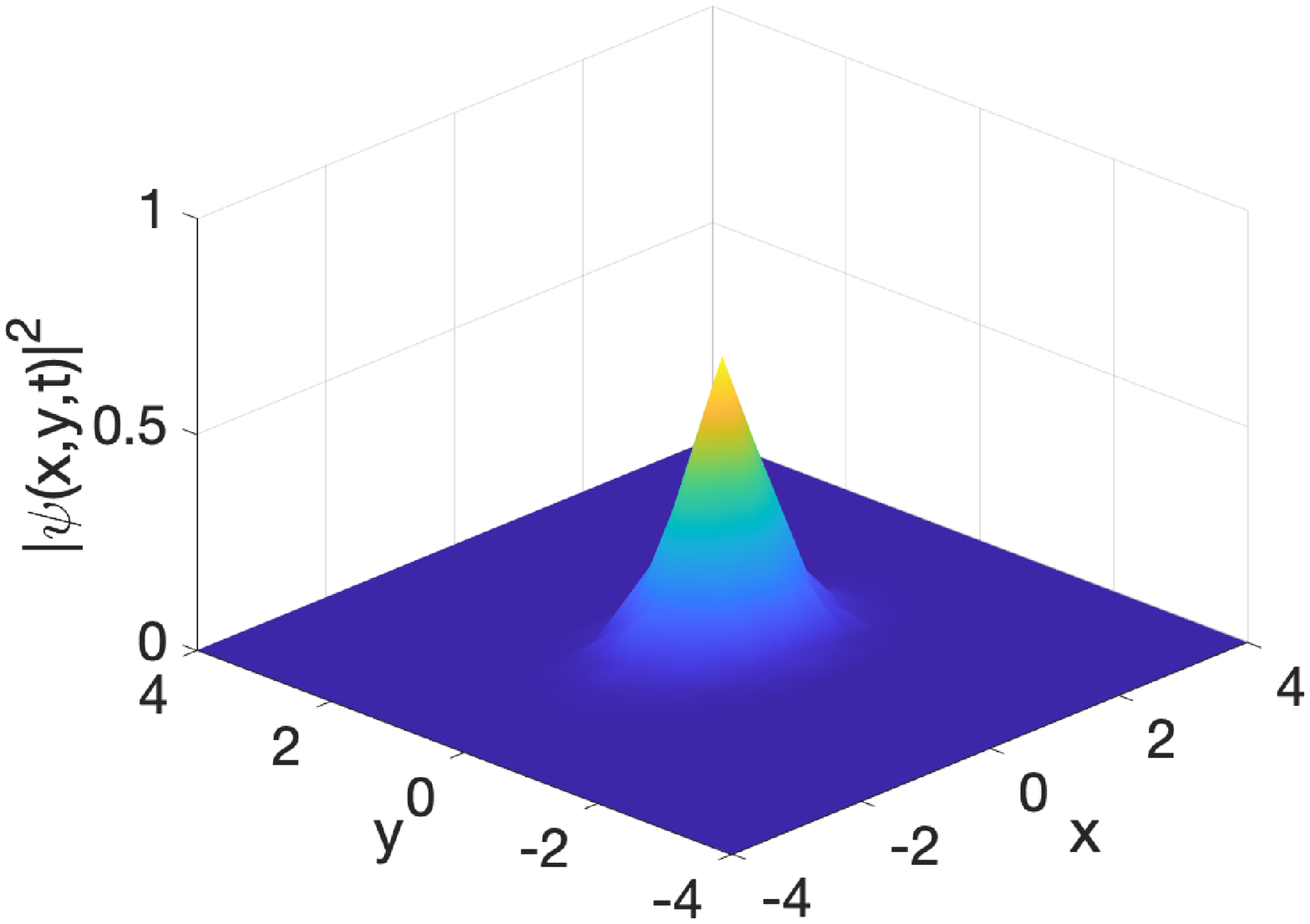}}
		\subfigure[]{\includegraphics[width=0.4200\textwidth]{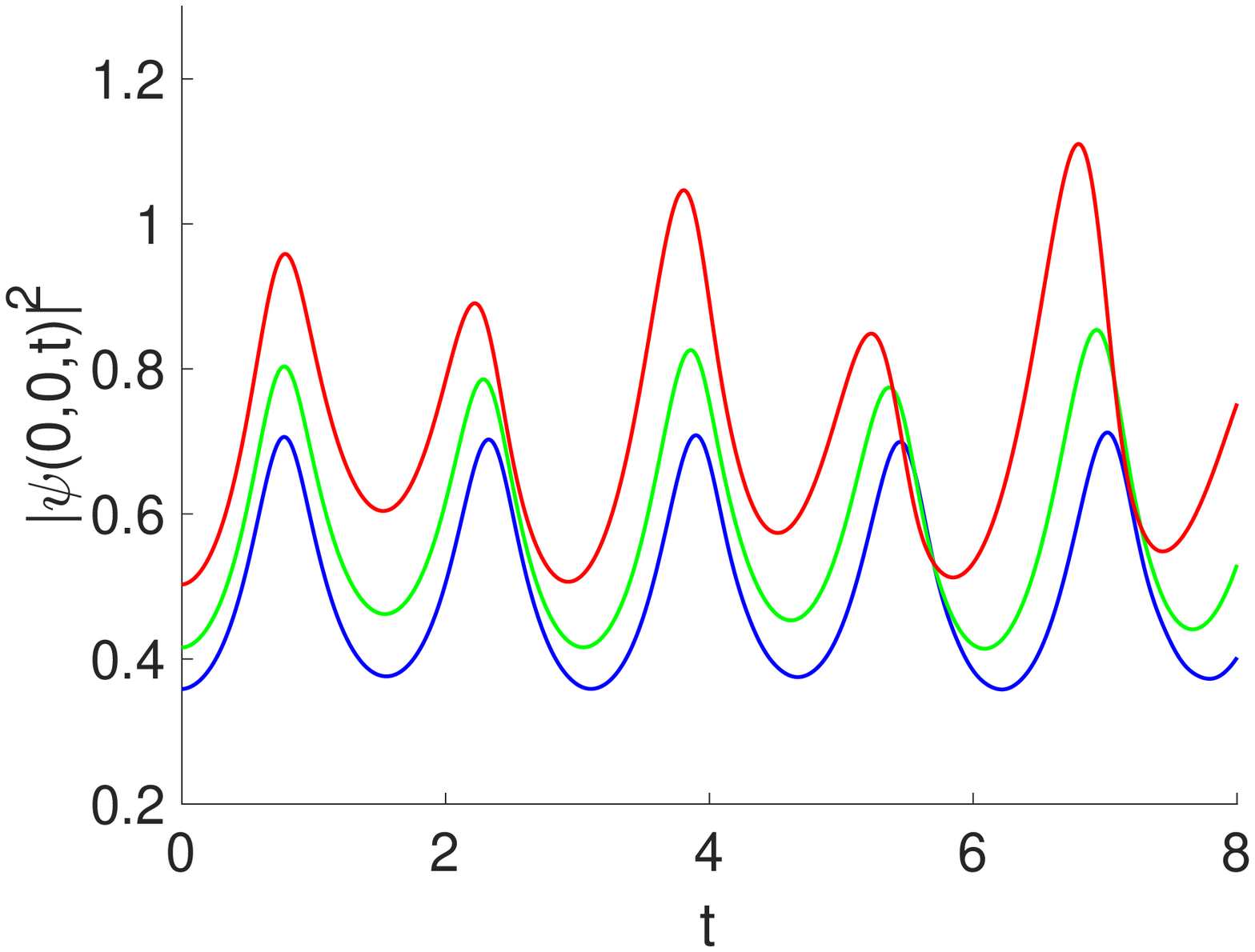}}
	}
	\caption{Same plots as shown in Figure 3, but for a condensate with an attractive interaction. (a) Surface plot of $|\psi(x,y,t)|^{2}$ at time $t=2.5$ and $g=-2$. Panel (b) shows $ |\psi(0,0,t)|^{2}$ for $ g = -1$ (blue), $-2$ (green) and $-3$ (red).   } 
\end{figure}

Figure 3a shows the density $|\psi(x,y,t)|^{2}$ for the anisotropic trap with $\lambda = 2$ and $g=5$. Panel 3b demonstrates the breathing frequency of this condensate depending on values of the nonlinear constant $g$. When $g$ increases, the initial density $|\psi(x,y,0)|^{2}$ at time $t=0$ lowers due to a repulsive interparticle interaction. However, it has been seen that the nature of all three curves is different from what has been appeared in Figure 2b. This may happen because the trap is contracted along the $y$-axis and particles could move higher and closer.

Figure 4 shows results for the system composed of the particle with a negative scattering length ($a_{s} <0$) which makes the system attractive ($g<0$). Panel 4a presents the density $|\psi(x,y,t)|^{2}$ for the anisotropic trap $g=-2$. As expected, similar patterns appear in Figure 4b like what was seen in Figure 2b since the anisotropic trap may support more particles in the center of the trap. We note that when an interaction becomes stronger (as $g$ gets larger/smaller value), uneven (up and down) nature appears (red curve in panels 3b and 4b). We also note that results for our 2D harmonic trap calculations are stable for the grid size parameters and the time interval we employed.

\section{Conclusion} 

In this paper we have solved the time-dependent Gross-Pitaevskii equation using split-step technique combined with the Legendre-pseudospectral method. From our numerical experiment for a non-linear Schr\"{o}dinger equation it's been shown that the split-step procedure is more accurate than an ordinary pseudospectral method. It has also been shown that the split-step Legendre-pseudospectral approach can be competitive one against the split-step Fourier pseudospectral approach due to the it's non-uniform grid point distribution along with the requirement of relatively few grid points. Furthermore, the SSLS approach has been tested for the one-dimensional and two-dimensional time- dependent GPE with harmonic trap to simulate the dynamics of the Bose-Einstein condensate with repulsive and attractive interactions. Making the expansion of the 1D harmonic trapping potential, the breathing modes of the condensate have been revealed as a function of the time for various values of the nonlinear constant of the condensate. Moreover, the simulation of those modes has been discussed for 2D harmonic trap which contracts from a looser one to a more confining trap.

\end{document}